\documentclass[aps,twocolumn,nofootinbib,showpacs,floatfix]{revtex4}
\usepackage{bm} 
\usepackage{amsmath}
\usepackage{amssymb}
\usepackage{bbm}
\usepackage{slashed}
\usepackage{graphicx}
\def\bea{\begin{eqnarray}}
\def\eea{\end{eqnarray}}

\begin{document}

\title{First results for electromagnetic three-nucleon form factors from high-precision two-nucleon interactions}

\author{S\'ergio Alexandre Pinto}
\affiliation{Centro de F\'isica Nuclear da Universidade de Lisboa, 1649-003 Lisboa, Portugal \\and Departamento de F\'isica da Universidade de \'Evora, 7000-671 \'Evora, Portugal}

\author{Alfred Stadler}
\affiliation{Centro de F\'isica Nuclear da Universidade de Lisboa, 1649-003 Lisboa, Portugal \\and Departamento de F\'isica da Universidade de \'Evora, 7000-671 \'Evora, Portugal}

\author{Franz Gross}
\affiliation{Thomas Jefferson National Accelerator Facility, Newport News, VA 23606\\ and College of William and Mary, Williamsburg, VA 23187}

\date{\today}

\begin{abstract}
The electromagnetic form factors of the three-nucleon bound states were calculated in Complete Impulse Approximation in the framework of the Covariant Spectator Theory for the new high-precision two-nucleon interaction models WJC-1 and WJC-2. The calculations use an approximation for the three-nucleon vertex functions with two nucleons off mass shell. 
%, which is tested and found to be very good. 
The form factors with WJC-2 are close to the ones obtained with the older model W16 and to nonrelativistic potential calculations with lowest-order relativistic corrections, while the form factors with the most precise two-nucleon model WJC-1 exhibit larger differences. These results can be understood when the effect of the different types of pion-nucleon coupling used in the various models is examined. 
\end{abstract}
\preprint{nucl-th/08????}

\maketitle

\section{Introduction}

The electromagnetic form factors of nuclei provide important information about their internal structure. They have been used extensively in order to test models of the nuclear dynamics and of the associated electromagnetic currents. As electron scattering experiments, such as the ones performed at Jefferson Lab, reach larger and larger values of the momentum transferred by a virtual photon to the struck nucleus, it becomes increasingly important to incorporate the requirements of special relativity in a reliable way into the theoretical description of the process.

The Covariant Spectator Theory (CST) \cite{CST} was designed as a manifestly covariant theory, especially suited for the description of few-nucleon problems. In a recent paper \cite{Pin09a}, we presented the first CST calculations of the electromagnetic three-nucleon ($3N$) form factors in Complete Impulse Approximation (CIA), which is defined as the complete CST $3N$ current \cite{Gro04} except for interaction currents, \textit{i.e.}, diagrams where the photon couples to an intermediate interacting two-nucleon ($NN$) system. However, the term ``impulse approximation'' can be misleading because it depends on the framework used. 

For instance, in a very successful approach used by the Pisa-Jlab collaboration \cite{Mar98, Mar05}, the dynamics is based on the nonrelativistic Schr\"odinger equation, and relativistic corrections are added perturbatively. We call the corresponding impulse approximation with relativistic corrections ``IARC''. In this framework, two- and three-body interaction currents are later added to the IARC results. These include first-order $\gamma\pi NN$ contact interactions which are equivalent to the already at the CIA level automatically---and to all orders---included ``Z-graphs'' in the CST. This is an example of a more general observation: what counts as interaction current in one approach may be part of the impulse approximation in another.

In Ref.~\cite{Pin09a}, our focus was to study the model dependence of the electromagnetic $3N$ form factors in CST. We performed calculations for a family of closely related relativistic two-nucleon interaction models and found that the CST results behave very reasonably. In most cases, a direct comparison of our CIA results with experimental data is not useful because we expect interaction currents to be significant. However, the comparison with IARC results is instructive, and it appears that the surprisingly close agreement between the two approaches (at least for $Q\alt 4$ fm$^{-1}$), when models are compared that yield the same $3N$ binding energy, is no coincidence. The main reason seems to be that all CST models used in the comparison employ pseudovector coupling for the pion-nucleon vertices. This kind of coupling suppresses negative-energy states (corresponding to Z-graphs), which are included in CIA but not in IARC. It is therefore understandable that no large differences between the two calculations emerge, as long as other aspects of the dynamics in the two approaches are comparable.

It would be interesting to submit this interpretation to a test. One only needs to perform two calculations in CIA with two $NN$ models that are as similar as possible in their ability to describe the $NN$ data and the $3N$ binding energy. One of them should be based on pure pseudovector pion-nucleon coupling, while the other should include an admixture of pseudoscalar pion-nucleon coupling and thus increase the weight of Z-graphs. If the above interpretation is correct, the model with pure pseudovector coupling will be close to the IARC result, while there should be larger deviations in the case of the model with some pseudoscalar coupling.

We are indeed in a position to perform this test. In a recent paper \cite{Gro08}, we published two realistic CST models for the neutron-proton interaction, both of which describe the $np$ scattering observables with $\chi^2/N_\mathrm{data} \sim 1$ for the most recent 2007 data base. The first model, WJC-1,  based on the exchange of 8 bosons and fitted with 27 adjustable parameters, features a mixture of pseudovector and pseudoscalar pion-nucleon coupling. The second model, WJC-2, based on the exchange of 6 mesons and with only 15 adjustable parameters, uses pure pseudovector pion-nucleon coupling. The two models can be considered to be essentially on-shell equivalent, and both reproduce also the experimental value of the triton binding energy of 8.48 MeV.

There is, however, one obstacle to performing the CIA calculations with models WJC-1 and WJC-2: Some of the diagrams that comprise the CIA $3N$ current depend on the $3N$ vertex function with \textit{two} nucleons off mass shell. A computer code for the calculation of these vertex functions for the new models WJC-1 and WJC-2 is at present in development, but not yet ready to be used in the calculation of the $3N$ form factors. 

This obstacle can be overcome if we apply an approximation in which $3N$ vertex functions with two nucleons off mass shell are appropriately replaced by vertex functions with only one nucleon off mass shell. 
%In this approximation, which we call CIA-0, the fully off-shell two-nucleon scattering amplitudes of WJC-1 and WJC-2 are no longer required. 
We can test the quality of this approximation, which we call ``CIA-0'', by applying it to one of the models previously used in \cite{Pin09a} and comparing the approximate form factors to the respective full CIA result. 

%Finding a good approximation to the full CIA which avoids all Feynman diagrams containing the full off-shell two-nucleon amplitudes would be of great practical value, since these diagrams are rather complicated and their numerical evaluation is a very time-consuming task.  
Of course, if it turns out that CIA-0 is a reliable approximation, the $3N$ form factors obtained from the realistic models WJC-1 and WJC-2 will be of high interest by themselves, not just as a means to evaluate the effect of Z-graphs. While the family of models used in \cite{Pin09a} gives a good description of the $NN$ data, they cannot compete in precision with our new models.

This paper is divided into four sections. After the introduction, Section \ref{sec:CIA-0} describes the $3N$ current and defines the CIA-0 approximation. Section \ref{sec:Results} presents and discusses the numerical results obtained, and in Section \ref{sec:Conclusions} we draw our conclusions.

\section{The $3N$ current in CIA and CIA-0}
\label{sec:CIA-0}

\begin{figure*}%fig4
\includegraphics[width=6.4cm,angle=-90]{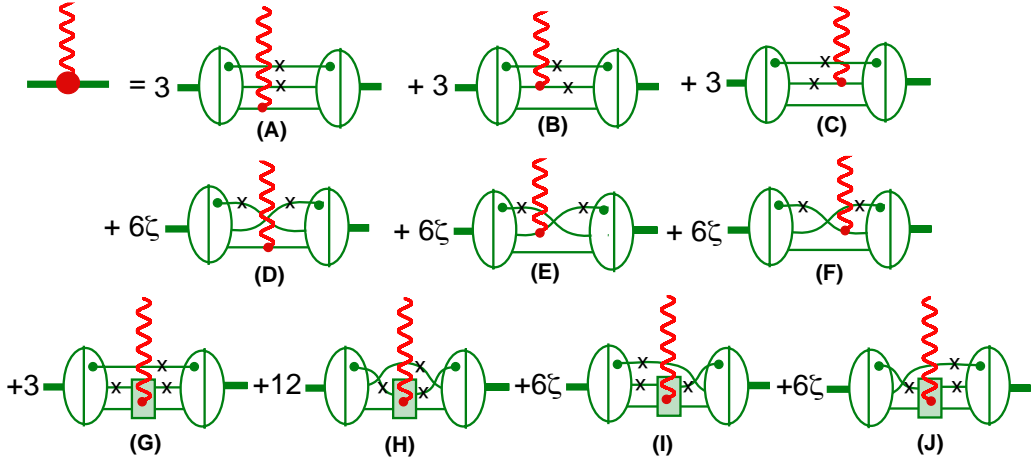}
\caption{(Color online) 
The electromagnetic $3N$ current in CST for elastic electron scattering from the $3N$ bound state.  A cross on a nucleon line indicates that the particle is on mass shell. Diagrams (A) to (F) define the complete impulse approximation (CIA), in which the photon couples to single nucleons, which can be off-shell (A and D) or on-shell before or after the photon-nucleon vertex (B, C, E, and F). The approximation denoted CIA-0 replaces the vertex function with two nucleons off mass shell in diagrams (B), (C), (E), and (F) by a vertex function with only one nucleon off mass shell.
The interaction diagrams (G) to (J) describe processes in which the photon couples to two-body currents associated with the two-nucleon kernel.}
\label{fig:coreFF}
\end{figure*}

The complete form of the electromagnetic $3N$ current in CST was derived in Ref.~\cite{Gro04}, and used in \cite{Pin09a} for the first time to calculate the $3N$ form factors in CIA. Figure \ref{fig:coreFF} displays the complete current, and CIA is defined through diagrams A to F.

We denote the photon four-momentum by $q$, and we label the nucleon four-momenta $k_i$ such that always $k_1^2=k_2^2=m^2$, where $m$ is the nucleon mass. For the cases where a nucleon absorbs a photon, we introduce the notation $k_i^\pm \equiv k_i \pm q$. The momentum $k_3$ is not an independent variable; in the CST, the energy-momentum four-vector is conserved and it is determined through the momenta of nucleons 1 and 2 and the total $3N$ momenta $P_t$ in the initial and $P'_t=P_t+q$ in the final state.

The $3N$ current in CIA is given in algebraic form by
\begin{widetext}
\begin{eqnarray}
J^\mu_{\rm CIA}&=&3e\int\!\!\int
\frac{m^2\,d^3k_1d^3k_2}{E(k_1)E(k_2)\,(2\pi)^6}
\sum_{\lambda_1\lambda_2}\Bigl\{
\bar\Psi_{\lambda_1\lambda_2\alpha'}(k_1,k_2;P'_t)\,
[1+2\,\zeta{\cal P}_{12}]
\,j_{\alpha'\alpha}^\mu(k^+_3,k_3)\,
\Psi_{\lambda_1\lambda_2\alpha}(k_1,k_2;P_t)\nonumber\\
& &+\bar\Gamma_{\lambda_1\beta'\alpha}(k_1,k^+_2;P'_t)
\,G_{\beta'\beta}(k^+_2)
\,j_{\beta\gamma}^\mu(k^+_2,k_2)\,u_{\gamma}(k_2,\lambda_2)\,
[1+2\,\zeta{\cal P}_{12}]\,
\Psi_{\lambda_1\lambda_2\alpha}(k_1,k_2;P_t)\nonumber\\
& &+\bar\Psi_{\lambda_1\lambda_2\alpha}(k_1,k_2;P'_t)\,
[1+2\,\zeta{\cal P}_{12}]\,\bar u_{\gamma}(k_2,\lambda_2)
\,j_{\gamma\beta'}^\mu(k_2,k^-_2)\,G_{\beta'\beta}(k^-_2)\,
\Gamma_{\lambda_1\beta\alpha}(k_1,k^-_2;P_t)
\Bigr\}\, , \label{eq:CIA}
\end{eqnarray}
\end{widetext}
where $E(k) = \sqrt{m^2+{\bf k}^2}$, ${\cal P}_{12}$ is a permutation operator which interchanges particles 1 and 2, $\zeta$ is a phase with $\zeta=+1(-1)$ for bosons (fermions), $G_{\beta'\beta}(k)$ is the propagator of an off-shell nucleon with four-momentum $k$, and $j_{\alpha'\alpha}(k',k)$ is the single nucleon current for
off-shell nucleons with incoming (outgoing) four-momentum
$k$ ($k'$). Summation over repeated Dirac indices is implied. The $3N$ vertex functions $\Gamma$ are solutions of Faddeev-type CST integral equations \cite{Sta97b} and were obtained numerically for the $NN$ interaction models considered here \cite{Sta97,Gro08}. The ``relativistic wave functions'' are defined as 
\begin{eqnarray}
\Psi_{\lambda_1\lambda_2\alpha}(k_1,k_2;P_t)=
G_{\alpha\alpha'}(k_3)
\Gamma_{\lambda_1\lambda_2\alpha'}(k_1,k_2;P_t)
\, ,
\label{wavefunction}
\end{eqnarray}
and we use a shorthand for the contraction of Dirac indices with nucleon helicity spinors with helicity $\lambda_i$,
\begin{multline}
 \Gamma_{\lambda_1\lambda_2\alpha'}(k_1,k_2;P_t) \equiv \\
\bar{u}_{\alpha_1}(k_1,\lambda_1) \bar{u}_{\alpha_2}(k_2,\lambda_2)
\Gamma_{\alpha_1\alpha_2\alpha'}(k_1,k_2;P_t) \, .
\end{multline}
In the second and third line of Eq.~(\ref{eq:CIA}), corresponding to diagrams (B+C+E+F) of Fig.~\ref{fig:coreFF}, the vertex function appears with \textit{two} nucleon momenta off mass shell. 
The solutions of the CST equation for the $3N$ bound state have only \textit{one} nucleon (nucleon 3, by convention) off mass shell, but one can obtain vertex functions with two off-shell particles through an iteration of the $3N$ equation with an off-shell two-nucleon scattering amplitude, 
\begin{eqnarray}
\lefteqn{
\Gamma_{\lambda_1\beta\alpha}(k_1,k^-_2;P_t)=-\int
\frac{m\,d^3k'_2}{E(k'_2)\,(2\pi)^3}
}\nonumber \\
&&
\times
\sum_{\lambda'_2}
M_{\beta\alpha,\lambda'_2\alpha'}(k^-_2,k_2';P_{23})
2\,\zeta\,{\cal P}_{12}\,
\Psi_{\lambda_1\lambda'_2\alpha'}(k_1,k'_2;P_t)
\, . \nonumber \\
\label{eq:3NFadoff}
\end{eqnarray}
Here, $M_{\beta\alpha,\lambda'_2\alpha'}(k^-_2,k_2';P_{23})$ is the scattering amplitude of nucleons 2 and 3 with total pair momentum $P_{23}$, and the final state momentum of nucleon 2, $k^-_2$, is off mass shell, while its initial-state momentum, $k'_2$, is on mass shell (nucleon 3 is off shell in either state).

As pointed out in the Introduction, the $3N$ vertex functions with both nucleons off-shell in the final state are not available at this time for the new $NN$ interactions WJC-1 and WJC-2. Moreover, it is rather awkward to calculate and manipulate these double-off-shell vertex functions numerically, because with one additional continuous variable (the off-shell energy of nucleon 2) they occupy much more computer storage space and slow down the calculations. 

For these practical reasons, we introduce here a simple approximation which replaces the vertex functions with two nucleons off-mass-shell by others with only one nucleon off mass shell.

In order to motivate this approximation, consider for instance diagrams (C) and (F) of Fig.\ \ref{fig:coreFF}. The vertex function in the initial state, $\Gamma_{\lambda_1\beta\alpha}(k_1,k^-_2;P_t)$, depends on the off-shell momentum $k_2^-=k_2-q$. Since $k_2$ is on mass shell, for small photon momenta $q$ the momentum $k_2^-$ is also almost on mass shell. We may therefore expand the vertex function in the off-shell energy of nucleon 2 around its on-shell value. If we keep only the zeroth-order term of the expansion and eliminate the corresponding negative-energy channel of nucleon 2 (with negative $\rho$-spin), we obtain a known vertex function with two nucleons on mass shell. We call this approximation ``CIA-0,'' referring to the zeroth-order expansion involved.

While this approximation is easy to apply, its formulation is somewhat awkward because of its frame-dependence. In our numerical calculations, the $3N$ vertex function is expressed in terms of variables for nucleons 2 and 3 which are defined in the rest frame of the (23) pair where the CST equation for the two-nucleon scattering amplitudes is solved numerically. We can write
\begin{equation}
\Gamma_{\lambda_1\beta\alpha}(k_1,k^-_2;P_t) = 
\Gamma_{\lambda_1\beta\alpha}(k_1,L(k^-_{23}) \tilde{k}^-_2;P_t) \, ,
\label{eq:vf}
\end{equation} 
where the Lorentz transformation $L(k^-_{23})$ takes the system of nucleons 2 and 3 from its rest frame, where their momenta are $\tilde{k}_2^-$ and $\tilde{k}_3$, to the $3N$ rest frame, where their momenta are $k^-_2=L(k^-_{23})\tilde{k}^-_2$ and $k_3=L(k^-_{23}) \tilde{k}_3$ and where their total two-body momentum is $k^-_{23}=k^-_2 + k_3=P_t-k_1$.

We define now the four-momentum $\tilde{r}_2^-$ to have the same three-vector part as $\tilde{k}_2^-$ but to be on mass shell, \textit{i.e.}, $\tilde{r}_2^-=(E(\tilde{k}_2^-),\tilde{\mathbf{k}}_2^-)$, and we replace the momentum $\tilde{k}^-_2$ by $\tilde{r}_2^-$ in the vertex function (\ref{eq:vf}).

In order to eliminate the negative energy states of nucleon 2, we first write the propagator of nucleon 2 in terms of its form in the pair rest frame,
\begin{multline}
  G_{\beta'\beta}(k_2^-) = 
S_{\beta'\beta_1}(L(k^-_{23})) \\
\times \left( \frac{m+\tilde{\slashed{k}}_2^-}{m^2-(\tilde{k}_2^-)^2-i\epsilon} \right)_{\beta_1\beta_2}
S^{-1}_{\beta_2\beta}(L(k^-_{23})) \, ,
\end{multline} 
where $S(L(k^-_{23}))$ is the Dirac space representation of the Lorentz transformation $L(k^-_{23})$.
%, which in our case includes apart from a boost also rotations \cite{Pin09a}.

Now we keep only the component with positive $\rho$-spin in the pair rest frame,
\begin{align}
% G(\tilde{k}_2^-) = &
\frac{m+\tilde{\slashed{k}}_2^-}{m^2-(\tilde{k}_2^-)^2-i\epsilon} \longrightarrow 
%\nonumber\\
%& 
\frac{m}{E(\tilde{k}_2^-)}\frac{\Lambda_+(\tilde{r}_2^-)}{\tilde{k}_{20}^--E(\tilde{k}_2^-)-i\epsilon} \, ,
\end{align} 
with the positive-energy projector
\begin{equation}
 \Lambda_+(\tilde{r}^-_2)=\frac{m+\tilde{\slashed{r}}_2^-}{2m} \, .
\end{equation} 
The approximation CIA-0 can then be defined as the replacement
\begin{multline}
G_{\beta' \beta}(k_2^-)
\Gamma_{\lambda_1\beta\alpha}(k_1,k^-_2;P_t)
 \longrightarrow \\
S_{\beta'\beta_1}(L(k^-_{23})) 
\frac{m}{E(\tilde{k}_2^-)}\frac{\left[ \Lambda_+(\tilde{r}_2^-)\right]_{\beta_1\beta_2}}{\tilde{k}_{20}^--E(\tilde{k}_2^-)-i\epsilon}
S^{-1}_{\beta_2\beta}(L(k^-_{23}))  \\
\times \Gamma_{\lambda_1\beta\alpha}(k_1,L(k^-_{23}) \tilde{r}_2^-;P_t)
\label{eq:CIA-0}
\end{multline}
in Eq.\ (\ref{eq:CIA}), as well as an analogous replacement for $\bar\Gamma_{\lambda_1\beta'\alpha}(k_1,k^+_2;P'_t) G_{\beta' \beta}(k_2^+)$, which occurs in diagrams (B) and (E).

Note that the projector $\Lambda_+$ eliminates negative-energy states of nucleon 2 in the two-body rest frame, but this does not eliminate all Z-graph contributions from the calculation. They are still present through the negative-energy states of nucleon 3, and they are also re-generated to some extent when the state of nucleon 2 is boosted to other frames. 

The approximation (\ref{eq:CIA-0}) may look complicated, but it is actually easy to implement in our numerical calculations. For instance, in the case of diagram C it merely amounts to replacing in Eq.\ (B64) of Ref.\ \cite{Pin09a} the off-shell energy $\tilde{p}^{0}$ of nucleon 2 by the corresponding on-shell value $E(\tilde{p})$ in the argument of the partial wave vertex function 
$C(  q \tilde{p}^{0} \tilde{p} M j m \lambda_{1} \lambda_{2} \lambda_{3}  \rho_{2} \rho_{3} T \mathcal{T}_{z}  )$, and restricting the summation over the $\rho$-spins of nucleon 2 to the positive-energy value $\rho_2=+$ only.

The electromagnetic current for an off-shell nucleon can be written in the form 
\begin{align}
j_N^\mu(k',k) = &f_0(k'^2,k^2)\,\,F_{1N}(Q^2)\, \gamma^\mu \nonumber\\
+&f'_0(k'^2,k^2)\,F_{2N}(Q^2)\,\frac{i \,\sigma^{\mu \nu}
q_\nu }{2m}  \nonumber\\
+& g_0(k'^2,k^2) F_{3N}(Q^2) \Lambda_- (k')  \gamma^\mu 
\Lambda_- (k)\,  ,  \label{eq:jmu}
\end{align}
where $f_0$, $f'_0$, and $g_0$ are nucleon off-shell form factors associated with the boson-nucleon vertices, and $F_{1N}$ and $F_{2N}$ are the usual electromagnetic Dirac and Pauli form factors. Since $\Lambda_-$ projects onto negative energy states, the form factor $F_{3N}$ belongs to a term that contributes only if the nucleon is in a negative-energy state before and after the photon-nucleon vertex. We adopt the usual convention $Q^2=-q^2$.

The isospin dependence of the electromagnetic form factors is, for $i=\{1,2,3\}$ and the nucleon isospin projection $\tau^3$,
\begin{align}
F_{iN}(Q^2)=F_{ip}(Q^2) \frac{1+\tau^3}{2}+F_{in}(Q^2) \frac{1-\tau^3}{2}\, .
\end{align}

In previous calculations \cite{Pin09a}, we found that the $3N$ form factors are quite insensitive to the inclusion and variations of the off-shell nucleon form factors. Therefore we employ in the calculations of this work the simpler on-shell nucleon current, with $f_0=f'_0=1$ and $g_0=0$. For the Dirac and Pauli form factors, we chose the parameterization of Galster \cite{Gal71}, in order to compare with IARC results provided to us by Marcucci \cite{Mar08pc} who used the same parameterization.

With the CIA-0 approximation in place, the electromagnetic $3N$ form factors are calculated numerically from the $3N$ vertex functions, which were obtained by solving the $3N$ CST equation in helicity partial wave form. The applied techniques are described in detail in Ref.~\cite{Pin09a}.

\section{Presentation and discussion of the results}
\label{sec:Results}

\begin{figure*}[tbh]
%\hspace{-4cm}
\includegraphics[height=20cm]{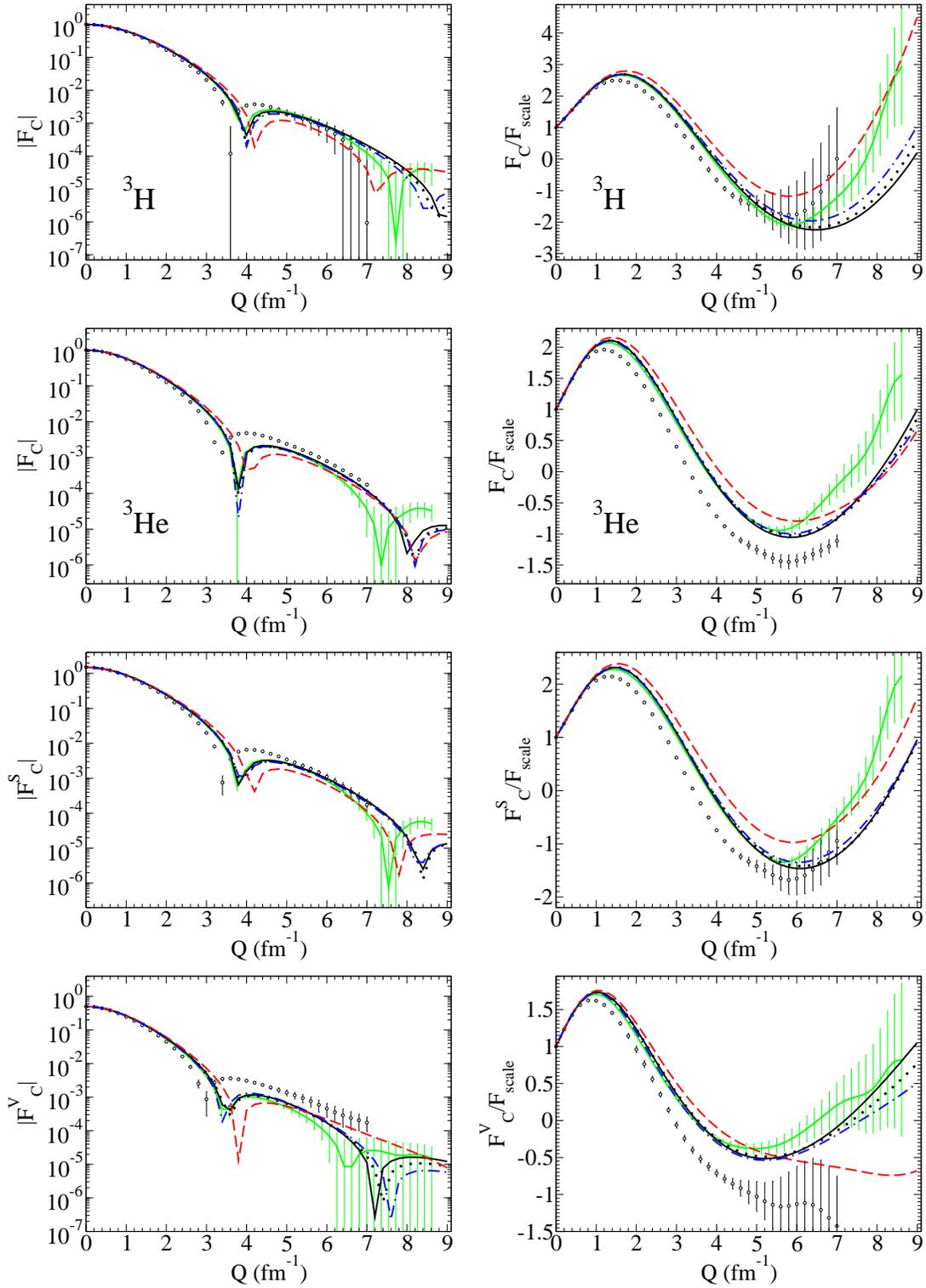}
\caption{(Color online) Charge form factors of the $3N$ bound states, $^3$H (first row), $^3$He (second row), and the isoscalar (third row) and isovector (fourth row) combinations. In each case, the figure on the left shows the form factor in the traditional semi-log plot, while the figure on the right shows the same form factor divided by a scaling function of Eq.~(\ref{eq:Fs}) \cite{Pin09a} on a linear scale. The solid line is the result for $NN$ model W16 in CIA, the dotted line is the approximation CIA-0 for the same model. The dashed line is model WJC-1, and the dash-dotted line is model WJC-2, both in CIA-0. For comparison, the solid line with theoretical error bars is the result of an IARC calculation by Marcucci \cite{Mar08pc} based on the AV18/UIX potential. All calculations employ the on-shell single-nucleon current, with the Galster parameterization of the nucleon form factors \cite{Gal71}. The full circles represent the experimental data %\cite{Sic01}.}
\cite{Col65,McCar77,Sza77,Arn78,Dun83,Ott85,Jus85,Bec87,Amr94}.}
\label{fig:chargeFF}
\end{figure*}

\begin{figure*}[htb]
%\hspace{-4cm}
\centering
  \includegraphics[height=20cm]{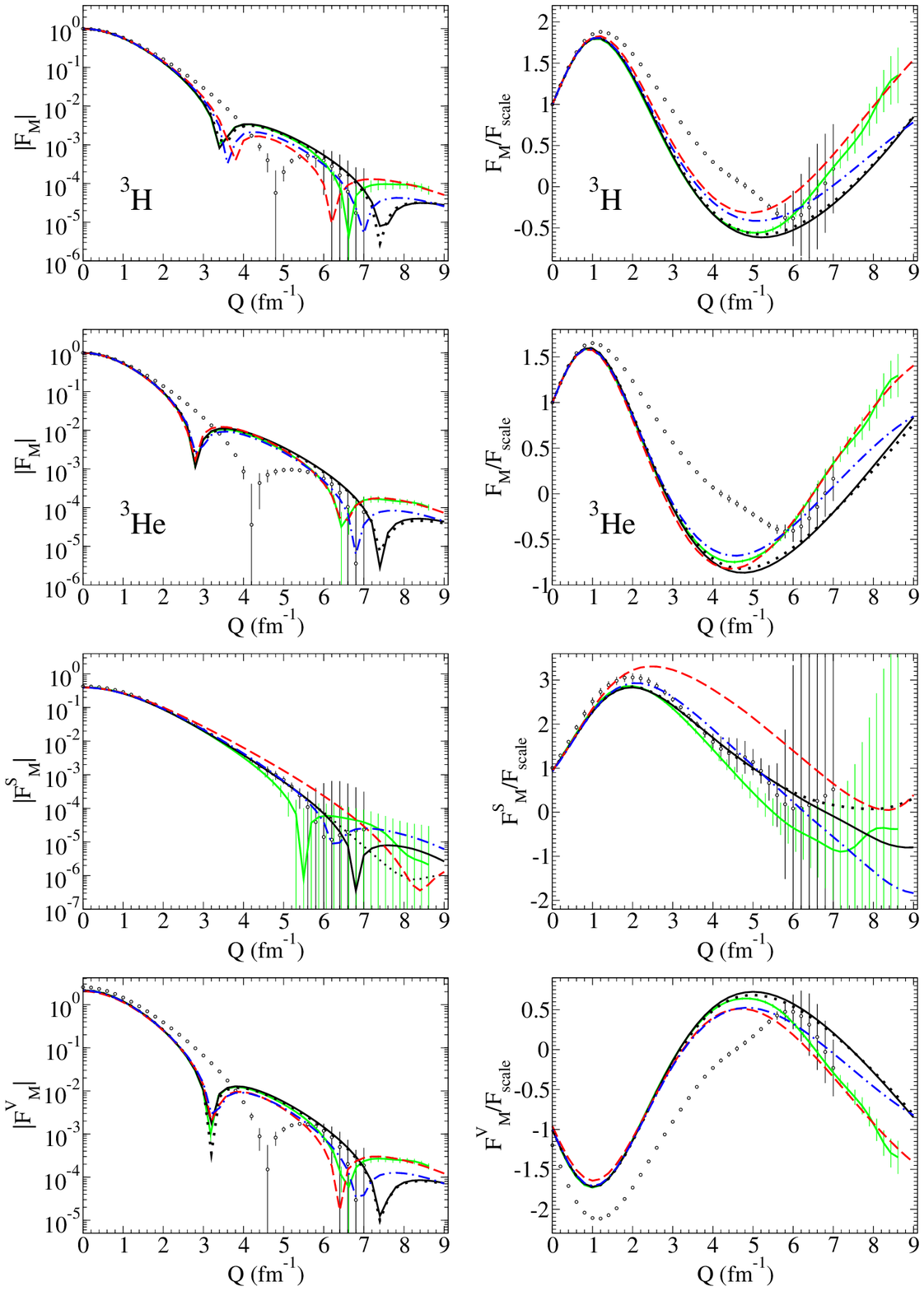}
\caption{(Color online) Magnetic form factors of the $3N$ bound states, $^3$H (first row), $^3$He (second row), and the isoscalar (third row) and isovector (fourth row) combinations. In each case, the figure on the left shows the form factor in the traditional semi-log plot, while the figure on the right shows the same form factor divided by a scaling function \cite{Pin09a} on a linear scale. The meaning of the various curves is the same as in Fig.~\ref{fig:chargeFF}.}
\label{fig:magFF}
\end{figure*}

We calculated the electromagnetic $3N$ form factors for three $NN$ interaction models, W16, WJC-1 and WJC-2, for momentum transfer up to $Q=9$ fm$^{-1}$.
The results are displayed in Figs.~\ref{fig:chargeFF} and \ref{fig:magFF}. 

Since the form factors fall several orders of magnitude, and the traditional log-plots tend to obscure differences in some places and overemphasize them in others, we divide them by simple scaling functions of the form 
\begin{equation}
F_s(Q)=F_0 e^{-Q/k}\, .
\label{eq:Fs}
\end{equation}
Table \ref{tab:Fs} shows the parameters of the scaling function for each case. We also list the magnetic moments in Table \ref{tab:magmom}, and the charge and magnetic radii in Table \ref{tab:rms}.

\begin{table}
 \caption{Parameters $F_0$ and $k$ (in fm$^{-1}$) of the scaling functions $F_s (Q)$ of Eq.~(\ref{eq:Fs}) by which the electromagnetic $3N$ form factors are divided in the figures with linear scale.}
\begin{tabular}{lcccc}
\hline \hline
& \multicolumn{2}{c}{charge f.f.} & \multicolumn{2}{c}{magnetic f.f.} \\
Form factor & $F_0$  & k  & $F_0$  & k \\
\hline
$^3$H  & 1 & 0.760488 & 1 & 0.871664\\
$^3$He  & 1 & 0.799411 & 1 & 0.912562\\
Isoscalar  & 1.5 & 0.778026 & 0.423 & 0.765425\\
Isovector  & 0.5 & 0.842695 & 2.13 & 0.889873\\
\hline\hline
 \end{tabular} 
\label{tab:Fs}
\end{table} 

First, we start with a comparison of the curves for W16 in CIA and in CIA-0, which clearly demonstrates the high quality of the approximation. The differences between the exact calculation and the approximation are hardly noticeable up to values of $Q$ around 7 fm$^{-1}$, and in general appear to be insignificant. We may therefore assume that the results for WJC-1 and WJC-2 obtained here only in CIA-0 should also be very close to the exact CIA result.

Note that there are caveats to this conclusion: the quality of CIA-0 compared to CIA was really tested only for W16, a model with a very smooth choice for the definition of the  kernel (and hence the vertex function) when both nucleons off-shell.  The WJC models have a more complex off-shell structure (corresponding to the prescription C discussed in Ref.~\cite{Gro08}) and their off-shell extrapolations will not be as smooth.  In addition, WJC-1 has a mixed pseudoscalar-pseudovector pion-nucleon coupling, and it is conceivable that the pseudoscalar part of this coupling might introduce further differences between CIA and CIA-0 to which W16 is not sensitive. 
%While one can have high confidence in the validity of CIA-0 for WJC-2, which has a very similar structure to W16, it is conceivable that the pseudoscalar part of the pion-nucleon coupling in WJC-1 spoils the quality of the approximation to some extent. 
Our conclusions must therefore be taken with these particular grains of salt. In any case, CIA-0 should be a better approximation to CIA at smaller $Q$, simply because the nucleon involved is taken less far off mass shell.

\begin{table}
\caption{Magnetic moments in nuclear magnetons (n.m.). The experimental values are from Ref.\ \cite{Til87}.}
\begin{tabular}{lcccc}
\hline \hline 
Model & $\mu (^3\mbox{H})$ & $\mu (^3\mbox{He})$ & $\mu_S$ & $\mu_V$ \\
\hline
%W16 (CIA) & 2.546 & -1.749 & 0.398 & -2.147  \\
W16 (CIA) & 2.544 & -1.747 & 0.400 & -2.144  \\
%W16 (CIA-0) & 2.545 & -1.747 & 0.399 & -2.146  \\
W16 (CIA-0) & 2.543 & -1.743 & 0.400 & -2.143  \\
%WJC-1 (CIA-0) & 2.497 & -1.564 & 0.466 & -2.030  \\
WJC-1 (CIA-0) & 2.441 & -1.648 & 0.396 & -2.044  \\
%WJC-2 (CIA-0) & 2.542 & -1.768 & 0.387 & -2.155  \\
WJC-2 (CIA-0) & 2.525 & -1.742 & 0.391 & -2.134  \\
IARC & 2.572 & -1.763 & 0.404 & -2.168\\
%IARC & 2.572 & -1.763 & 0.404 & -2.168\\
\hline
Experiment & 2.979 & -2.128 & 0.426 & -2.553 \\
\hline\hline
\end{tabular}
\label{tab:magmom} 
 \end{table}

We turn now to a comparison of the form factors for different models of the $NN$ interaction.
The figures show that the WJC-2 form factors stay close to the ones of W16, while, in most cases, WJC-1 begins to deviate somewhat already at smaller values of Q. 
WJC-2 and W16 are also close to the IARC results, typically up to about $Q=6$ fm$^{-1}$. This supports the conjecture made in the Introduction, namely that the suppression of Z-graphs through the use of pseudovector pion-nucleon coupling is mainly responsible for the close agreement between the CST and IARC. 

Apart from the issue of the type of pion-nucleon coupling, the CST models include other boson-exchanges with off-shell coupling. Most notably, those due to scalar isoscalar ($\sigma_0$) and isovector ($\sigma_1$) exchanges have been found to have a very strong influence on the quality of the $NN$ fits and on the triton binding \cite{Sta97}. One might expect them to have a strong influence on the $3N$ form factors as well.

The results indicate that this is only indirectly the case, namely through their effect on the binding energy. When the scalar off-shell coupling strength is varied without constraining the triton binding energy, the $3N$ form factors show substantial variations \cite{Pin09a}. 
On the other hand, models W16 and WJC-2 have quite different scalar off-shell coupling constants, but yield the same triton binding energy. The close similarity of the $3N$ form factors, at least up to intermediate values of $Q$, implies that the electromagnetic structure of the $3N$ bound state is not modified too much by the scalar off-shell coupling. This conclusion receives even stronger support from the observation that also the IARC calculation, which of course has no off-shell couplings at all, essentially coincides with both W16 and WJC-2 up to about $Q=6$ fm$^{-1}$ in the charge form factors, and up to somewhat smaller values of $Q$ for the magnetic form factors.

It follows then that most of the different behavior of the WJC-1 form factors cannot be attributed to the models larger scalar off-shell couplings \cite{Gro08}, unless the dependence turns out to be highly non-linear.
 
What about the contributions of Z-diagrams and pion exchange currents?  In Ref.~\cite{Mar98} it was shown that pion exchange currents bring the calculations closer to the experimental data.  The form of these exchange currents depends on the structure of the $\pi NN$ coupling.  For pseudoscalar coupling there is no $\gamma\pi NN$ contact interaction, but there are large contributions from Z-diagrams.  The opposite is true for pseudovector coupling.  Here, the minimal substitution $q_\pi\to q_\pi-eA$ into the momentum dependent interaction, $g_{\pi NN}\gamma^5\slashed{q}_\pi/2m$  (where $q_\pi$ is the pion momentum), leads to the contact interaction, $eg_{\pi NN}\gamma^5\gamma^\mu/2m$, but the Z-diagrams that are produced by a pseudovector interaction are strongly suppressed (and vanish in the nonrelativistic limit). Furthermore, there is an equivalence theorem that has been know for many decades \cite{Ris72}: in the nonrelativistic limit, the Z-diagrams derived from pseudoscalar coupling are identical to the contact interaction derived from pseudovector coupling (and the pseudovector Z-diagrams vanish).  The pseudovector contact interaction equals the pseudoscalar Z-diagram.  

\begin{table}
 \caption{Root-mean-square charge and magnetic radii in fm. The experimental values are from Ref.\ \cite{Amr94}.}
\begin{tabular}{lcccc}
\hline \hline 
Model &  
$r_\mathrm{ch}(^3\mbox{H})$ & $r_\mathrm{ch}(^3\mbox{He})$ & $r_\mathrm{mag}(^3\mbox{H})$ & $r_\mathrm{mag}(^3\mbox{He})$\\
\hline
%W16 (CIA) &  1.718 & 1.900 & 1.928 & 2.031 \\
W16 (CIA) &  1.718 & 1.900 & 1.915 & 2.001 \\
%W16 (CIA-0) &  1.720 & 1.900 & 1.927 & 2.031 \\
W16 (CIA-0) &  1.720 & 1.901 & 1.915 & 2.001 \\
%WJC-1 (CIA-0) &  1.700 & 1.879 & 2.120 & 1.429 \\
WJC-1 (CIA-0) &  1.700 & 1.879 & 1.901 & 2.035 \\
%WJC-2 (CIA-0) &  1.722 & 1.903 & 1.974 & 2.166 \\
WJC-2 (CIA-0) &  1.722 & 1.904 & 1.904 & 2.027 \\
\hline
Experiment & 1.755  & 1.959  & 1.840  & 1.965  \\
          & $\pm 0.086$ & $\pm 0.030$ & $\pm 0.181$ & $\pm 0.153$ \\
\hline\hline
\end{tabular}
\label{tab:rms} 
\end{table} 

This discussion is helpful in interpreting the difference between our results for WJC-1 and WJC-2/W16/IARC.  The CIA calculations reported here include Z-diagrams (to all orders) but do not include contact interactions.  These are included in the diagrams G-J of Fig.~\ref{fig:coreFF}, and are excluded from both the CIA and CIA-0 calculations.  They must be added separately, just as in the work of Ref.~\cite{Mar98}.
Now, the pion-nucleon vertex can be written in the general form $g_{\pi NN}[\lambda \gamma^5 + (1-\lambda)\gamma^5 \slashed{q}_\pi/2m]$, where $\lambda$ is the pseudoscalar-pseudovector mixing parameter \cite{Gro08}. The $NN$ models used in Ref.~\cite{Pin09a} such as W16, as well as WJC-2, use pure pseudovector coupling, with $\lambda=0$. Hence the Z-diagram contributions of these models are very small, and it is not surprising that they are quite close to the IARC result.  However, in WJC-1 the neutral and charged pions are treated separately, and the mixing parameter for the charged pions is $\lambda_{\pi^\pm}=-0.312$.  The Z-diagram contributions from this model should be large, but {\it of the opposite sign\/} from those from a pure pseudoscalar theory (corresponding to $\lambda_{\pi^\pm}=+1$).  Hence we can expect the Z-diagram contributions from WJC-1 to move the theory further away from the data, which is what we observe.  We expect this effect to be more than cancelled once the contact interactions are included, which for WJC-1 will have a strength 1.312 times a pure pseudovector coupling.

It is certainly not possible to draw very strong conclusions based on these results alone. There are simply too many variables in play, and it would require many more test calculations to try to disentangle them. 
However, it is a very interesting situation to have essentially on-shell equivalent interactions, which even agree in the $3N$ binding energy, but lead to different $3N$ form factors. It has often been argued that electromagnetic probes provide a means to distinguish otherwise equivalent $NN$ interaction models, and in this case we can actually see it happening. From this point of view, it is perhaps less surprising that WJC-1 differs somewhat from the other models, but rather that WJC-2, W16, and IARC are so close to each other. After all, IARC is calculated from the largely phenomenological nonrelativistic Argonne AV18 two-nucleon and UIX irreducible $3N$ force, where the latter is needed to make up for the missing $3N$ binding energy of the AV18 potential alone. In contrast, the CST models do not add any irreducible $3N$ forces, and it is through off-shell couplings---purely relativistic effects---that effective $3N$ forces are implicitly generated. Moreover, relativity is implemented in very different ways. It is not at all obvious that the two approaches should yield so similar results.

It would be premature to favor one or the other of the models WJC-1 or WJC-2 at this time. Their ability to reproduce the data can only be judged rigorously after all---or at least the dominant---interaction currents are added. Compared to the interaction currents of Ref.~\cite{Mar98}, some are already accounted for in CIA, others need to be added as well, such as the boson-in-flight terms. Probably more important than the latter, there are interaction currents induced by total-momentum dependencies in the vertices due to off-shell boson-nucleon coupling, which have never been evaluated. It may very well turn out that they have a stronger effect with WJC-1 than with WJC-2.

\section{Conclusions}
\label{sec:Conclusions}
We have performed the first calculations of the electromagnetic $3N$ form factors with the new covariant two-nucleon interaction models WJC-1 and WJC-2, which yield an excellent description of the neutron-proton observables below 350 MeV for the most recent 2007 data base \cite{Gro08}. The form factors were calculated in Complete Impulse Approximation (CIA), in which---for practical reasons---we replaced $3N$ vertex functions with two off-mass-shell nucleons by corresponding vertex functions with only one nucleon off mass shell. This procedure of approximating the full CIA results, denoted as ``CIA-0'', was tested with the older two-nucleon model W16, for which the full CIA result is also available, and found to be of very good quality.

We compare the form factors of WJC-1 and WJC-2 to those of W16, and also to calculations of nonrelativistic impulse approximation with relativistic corrections (IARC) by Marcucci and collaborators \cite{Mar08pc,Mar98}. Relating the observed differences in the various results to the underlying nuclear dynamics, we reach the following principal conclusions:

(i) The $3N$ binding energy determines the electromagnetic $3N$ form factors in impulse approximation up to unexpectedly large values of the transferred momentum. The closely related family of models investigated in Ref.~\cite{Pin09a}, where variations in the strengths of the scalar off-shell coupling led to significantly different $3N$ binding energies, showed much larger changes in the form factors than the models considered here, which all have the same binding energy.

(ii) The scalar off-shell coupling does not directly exert strong influence on the shape of the form factors. When the $3N$ binding energy is constrained to be equal, different scalar off-shell coupling strength can yield very similar form factors, as one can see comparing WJC-2 and W16.

(iii) In some cases, model WJC-1 deviates moderately from the others. This appears to be due to its mixed pseudoscalar-pseudovector pion-nucleon coupling (WJC-2 and W16 have pure pseudovector pion-nucleon coupling, whereas in the nonrelativistic framework of IARC the two couplings are equivalent). In CST, pseudoscalar pion-nucleon coupling automatically includes Z-diagrams, while they are suppressed for pseudovector coupling. When Z-diagrams are effectively added to IARC in the form of $\gamma\pi NN$ contact interactions \cite{Mar98}, the calculated form factors move closer to the experimental data, whereas the WJC-1 form factors lie further away than the models with pure pseudovector coupling. This is consistent, because the sign of the pseudoscalar coupling in WJC-1 is opposite to the one used in Ref.~\cite{Mar98}.

(iv) The results of this work confirm the conjecture formulated in the Introduction, namely that the reason for the good agreement of the CST models with the IARC results is the the suppression of Z-diagrams through pseudovector pion-nucleon coupling.

(v) The CST two-nucleon interaction models WJC-1 and WJC-2 not only give an excellent fit to the available two-nucleon scattering observables, but also provide a solid basis for a relativistic theory of the $3N$ system. Without additional irreducible $3N$ forces, the $3N$ binding energy is reproduced, and the electromagnetic $3N$ form factors turn out very similar to previous nonrelativistic results. No unusually large interaction currents seem to be required in order to achieve a quantitative description of the experimental data.

\begin{acknowledgments}
We thank L.\ Marcucci for providing the results of the IARC calculations.
S.\ A.\ P.\ and A.\ S.\ received support from FEDER and FCT under grant Nos.\ SFRH/BD/8432/2002 and POCTI/ISFL/2/275. F.\ G.\ was supported by Jefferson Science Associates, LLC under U.S. DOE Contract No.~DE-AC05-06OR23177.  A.\ S.\ thanks the Jefferson Lab Theory Group for its hospitality.
\end{acknowledgments}

\bibliographystyle{h-physrev3} % Choose Phys. Rev. style for bibliography
\bibliography{FBpapers}

\end{document}